\documentclass[aps,twocolumn]{revtex4}

\usepackage{graphicx}

\usepackage{dcolumn}
\usepackage{amsmath}
\usepackage{latexsym}
\usepackage{MnSymbol}
\usepackage{textcomp}
\makeatletter
\def\NAT@def@citea{\def\@citea{\NAT@separator}}
\makeatother
\usepackage{bm}

\begin{document}
\title{Percolative switching in transition metal dichalcogenide field-effect transistors at room temperature}

\author{Tathagata Paul,$^1$\footnotemark[3] Subhamoy Ghatak,$^1$\footnote{Current address: Institute of Scientific and Industrial Research, Osaka University, Ibaraki, Osaka 567-0047, Japan} and Arindam Ghosh$^1$\footnote[3]{e-mail:tathagata@physics.iisc.ernet.in, arindam@physics.iisc.ernet.in}}

\address{$^1$Department of Physics, Indian Institute of Science, Bangalore 560012, India}

\begin{abstract}
We have addressed the microscopic transport mechanism at the switching or ``on-off'' transition in transition metal dichalcogenide (TMDC) field-effect transistors (FET), which has been a controversial topic in TMDC electronics, especially at room temperature. With simultaneous measurement of channel conductivity and its slow time-dependent fluctuation (or noise) in ultra-thin WSe$_2$ and MoS$_2$ FETs on insulating SiO$_2$ substrates, where noise arises from McWhorter-type carrier number fluctuations, we establish that the switching in conventional backgated TMDC FETs is a classical percolation transition in a medium of inhomogeneous carrier density distribution. From the experimentally observed exponents in the scaling of noise magnitude with conductivity, we observe unambiguous signatures of percolation in random resistor network, particularly in WSe$_2$ FETs close to switching, which crosses over to continuum percolation at a higher doping level. We demonstrate a powerful experimental probe to the microscopic nature of near-threshold electrical transport in TMDC FETs, irrespective of the material detail, device geometry or carrier mobility, which can be extended to other classes of 2D material-based devices as well.
\end{abstract}

\maketitle
\begin{table*}[t]
\centering
\renewcommand{\arraystretch}{1.3}
\caption{Channel Length ($L$), Channel Width ($W$), details of electrical contact and layer number for the WSe$_2$ FET devices}
\label{tab:example}
\centering
\begin{tabular}{c|c|c|c|c}
    \hline
    Device Name & $L$ ($\mu$m) & $W$ ($\mu$m) & Contact Details & No. of layers\\
    \hline
    \hline

    WSeML\_2 & 2.8 & 3 & Ti/Au 10nm/50nm & 5\\
    \hline

    WSeBL & 1.5 & 5.3 & Ti/Au 10nm/50nm & 2\\
    \hline

    WSeML\_3 & 1.8 & 3 & Ti/Au 10nm/50nm & 5\\
    \hline

    MoSSL & 3 & 4 & Cr/Au 10nm/50nm & 1\\
    \hline

    MoSSL\_1 & 0.876 & 0.861 & Cr/Au 10nm/50nm & 1\\
    \hline

    MoSSL\_2 & 0.562 & 1.023 & Cr/Au 10nm/50nm & 1\\
    \hline
\end{tabular}
\end{table*}

The expanding family of atomically thin layers of semiconducting TMDC materials for electronic~\cite{Andras_Kis_MoS2_transistor,Peide_Ye_Channel_length_scaling,Andras_Kis_logic,Xiangfeng_Duan_logic,Subhamoy_ACS_Nano,Pradhan_ACS_Nano_2014,Andras_Kis_Metal_Insulator}, optoelectronic~\cite{Andras_Kis_MoS2_photodetectors_Nat_Nano,Kallol_Gr_MoS2_hybrid_Nat_Nano,Kallol_Solid_State_Communications,Pablo_WSe2_pn_junc_Nat_Nano,Mueller_WSe2_solar_energy_conversion_Nat_Nano,Xiaodong_Xu_WSe2_Led,Philip_Kim_pn_junction_heterostructure},~ valleytronic~\cite{K_F_Mak_Valley_Hall_Effect,Tony_F_Heinz_Valley,Xiaodong_Xu_Valley,Dong_Sun_Valley,Kim_Valley_Excitons_WSe2} and even piezo-electronic~\cite{Wenzhuo_Wu_Piezoelectricity_MoS2} applications, has defied a generic framework of electron transport because of diverse material quality, channel thickness dependent band structure, dielectric environment, and nature of substrates. A wide variety of physical phenomena ranging from variable range hopping~\cite{Subhamoy_ACS_Nano}, metal-insulator transition~\cite{Andras_Kis_Metal_Insulator} to classical percolative charge flow through inhomogeneous medium~\cite{C_T_Liang_disordered_MoS2,Ning_Wang_MIT_MoS2} were reported for MoS$_2$, the implications of which often lead to a conflicting microscopic scenario. At low carrier density, for example, hopping via single particle states trapped at short-range background potential fluctuations ($\sim$few lattice constants~\cite{Mott_Semiconductor_Book}) is incompatible to the observation of classical percolative conduction that requires long-range inhomogeneity in charge distribution, created when linear screening of the underlying charge disorder breaks down~\cite{Yigal_Meir_Percolation,S_Das_Sarma_Metal_Insulator_Transition,Si_inversion_layer_Shankar_Das_Sarma,Graphene_Nanoribbon_Shankar}. Observations of metal-insulator transition~\cite{Andras_Kis_Metal_Insulator,Ning_Wang_MIT_MoS2} have added to this controversy with both many-body Coulomb interaction as well as classical percolation transition being argued as possible driving mechanisms. With the emergence of new TMDC-based FETs, in particular WSe$_2$~\cite{Javey_Ali_WSe2_p_FETs,Kaustav_Banerjee_contacts_WSe2,E_Bucher_mobility,Saptarshi_Das_WSe2,Xiangfeng_Duan_Large_Area_p_WSe2,Lain_Jong_Li_Large_area_WSe2,Graphene_contact_WSe2,Kim_Valley_Excitons_WSe2,Andras_Kis_mobility}, WS$_2$~\cite{Artem_Mishchenko,Andras_Kis_WS2_anneal,Photosensor_WS2_Mauricio,WS2_light_emitting_transistors_Sanghyun}, and MoTe$_2$~\cite{Pradhan_ACS_Nano_2014}, the importance of a generic probe to the microscopic details of density (or gate voltage) dependent charge distribution that affect the mobility, subthreshold slope and other performance markers in TMDC electronics, is paramount.

Unlike variable range hopping in the strongly localized regime, however, identifying charge percolation in FET channels is less straightforward. Theoretically, classical percolation-limited conductivity in metal-insulator composites is characterized by established critical exponents~\cite{Kiss_Percolation_Noise,Yigal_Meir_Percolation,S_Das_Sarma_Metal_Insulator_Transition,C_T_Liang_disordered_MoS2,S_H_Kogan_Noise,Tremblay_Percolation_Exponents,1_by_f_noise_Webb,Sen_Percolation}, but the experimental difficulty lies in accurately determining the fraction $p$ of the conducting region, or ``puddles'', embedded inside the insulating matrix. Hence despite compelling evidence of long range inhomogeneity in the charge distribution in MoS$_2$ FETs~\cite{C_T_Liang_disordered_MoS2,Ning_Wang_MIT_MoS2}, its manifestation in transport remains indirect and confined only to low temperatures. A way to circumvent this difficulty involves measuring the low-frequency noise, or $1/f$-noise, in the channel conductivity $\sigma$, which also scales with $p$ with independent characteristic scaling exponents~\cite{Tremblay_Percolation_Exponents,Kiss_Percolation_Noise}, and diverges at the percolation threshold ($p_c$)~\cite{S_H_Kogan_Noise}. A direct relation between the normalized noise magnitude $N_\sigma$ and $\sigma$ thus eliminates the necessity to know either $p$ or $p_c$ accurately. In such a case,
\begin{equation}
\label{percolation}
N_\sigma=\frac{\langle\delta\sigma^2\rangle}{\sigma^2} \propto \sigma^{-\nu},
\end{equation}

\noindent where the scaling exponent $\nu$ assumes universal values depending on the nature of percolation. In the 2D lattice models of random resistor network, for example, $\nu \approx 0.86$ or $1.5$,~[REF:~\cite{Kiss_Percolation_Noise,Stauffer_Percolation_Theory}] whereas in the continuum percolation framework, such as the ``swisscheese model''~\cite{Sen_Percolation}, $\nu \approx 3.2$ for random void (RV, insulating voids in a conducting background) and $\approx 0.87$ for inverted random void (IRV, weakly connected conducting regions in an insulating matrix), respectively~\cite{S_H_Kogan_Noise,Tremblay_Percolation_Exponents}. In this work we have investigated the switching process in TMDC FETs by simultaneously measuring $N_\sigma$ and $\sigma$ as the electrical transport is tuned from the strongly insulating to the quasi-metallic regime using a gate voltage. While noise in all devices was found to primarily originate from carrier number fluctuations at the channel-substrate interface, the key result is the observation of an unambiguous scaling of $N_\sigma$ with $\sigma$, suggesting classical percolation in spatially inhomogeneous medium at the onset of conduction. Similar behavior for both WSe$_2$ and MoS$_2$ FETs imply that percolative switching could be generic to backgated TMDC FETs irrespective of material and device-related parameters.

We chose WSe$_2$ FETs as the primary experimental platform due to the following reasons: First, WSe$_2$ is an emerging TMDC FET material with several desirable properties ranging from high carrier mobility due to low effective mass of the carriers, ambipolar conduction and superior chemical stability compared to sulphides~\cite{Mueller_WSe2_solar_energy_conversion_Nat_Nano,Xiaodong_Xu_WSe2_Led,Andras_Kis_mobility,Javey_Ali_WSe2_p_FETs,Xiangfeng_Duan_Large_Area_p_WSe2,Lain_Jong_Li_Large_area_WSe2,Andras_Kis_WS2_anneal}. Second, in spite of the progress in standard electrical transport properties~\cite{Javey_Ali_WSe2_p_FETs,Kaustav_Banerjee_contacts_WSe2,E_Bucher_mobility,Saptarshi_Das_WSe2,Xiangfeng_Duan_Large_Area_p_WSe2,Lain_Jong_Li_Large_area_WSe2,Graphene_contact_WSe2,Kim_Valley_Excitons_WSe2,Andras_Kis_mobility}, the origin and magnitude of intrinsic $1/f$-noise, a crucial performance limiting factor in electronic device applications, in WSe$_2$ FETs is not known, and third, given the recent studies of noise in MoS$_2$ FETs~\cite{Kaustuv_Banerjee_MoS2_noise,Subhamoy_APL_Mat,Balandin_MoS2_noise,Junhong_Na_MoS2_noise,MoS2_noise_D_Sharma}, similar studies in WSe$_2$ allows identification of the generic aspects of noise processes in TMDC FETs, which in turn, provides crucial insight to microscopic details of charge distribution and disorder.

The experiments were carried out in FETs fabricated from ultra-thin films of  WSe$_2$ (and MoS$_2$) obtained via mechanical exfoliation on $p^{++}$-Si/(285~nm)SiO$_2$ substrates in the conventional backgated geometry (schematic and typical device image in Fig.~1a) (Details of the measurement technique provided in Supplementary Information). The details of the devices can be found in Table~I. In order to verify that the structural integrity of the WSe$_2$ channels were unaffected by the device fabrication process, we carried out Raman spectroscopy on, (1) original block of WSe$_2$ which was subsequently mechanically exfoliated to form the device WSeML\_2, (2) channel region prior to device fabrication, and (3) the same after device fabrication. The normalized Raman spectra (Fig.~1b) remains essentially unchanged, with characteristic Raman modes for multilayer WSe$_2$ including those at
$\approx 247.3$~cm$^{-1}$ ($E\textquotesingle^{1}$), $\approx 249.5$~cm$^{-1}$ ($A\textquotesingle_{1}$), and $\approx 308.5$~cm$^{-1}$ ($A_{1g}^2/A\textquotesingle_{1}^2$)~\cite{Terrones_Raman_MoS2}, implying no detectable structural damage due to the exfoliation or the device fabrication processes. Similar characterization for the MoS$_2$ channels and detailed surface characterization with atomic force microscopy can be found in the supplementary material. Measurement of both standard transport and noise were performed in a two-probe geometry via ac lock-in amplifier technique, where the latter was recorded by digitizing the amplifier output by a analog-to-digital converter over a bandwidth of $\approx 0.01 - 10$~Hz~\cite{Subhamoy_APL_Mat}.

In TMDC FETs, thermal annealing and environmental exposure have drastic effects on electrical mobility, threshold voltage and other parameters~\cite{Andras_Kis_mobility,Andras_Kis_WS2_anneal,Pablo_Bilayer_MoS2}. Recently, we have shown~\cite{Subhamoy_APL_Mat} that thermal annealing can also reduce the noise magnitude in MoS$_2$ FETs by over two orders of magnitude, through improved transparency of the electrical contacts~\cite{Andras_Kis_WS2_anneal,Pablo_Bilayer_MoS2}. We found a strong effect of annealing on the electrical transport in our WSe$_2$-FETs. In Fig.~1c-f, we illustrate the effect of annealing on both transport and noise characteristics in WSe$_2$ FETs, as well as the sensitivity of these parameters to environmental contamination. The inset of Fig.~1c shows the dependence of the source-drain current ($I_{sd}$) on gate voltage $V_g$ in device WSeML\_2 under three conditions: (a) in air/atmosphere prior to annealing (labelled as Air), (b) in vacuum ($\approx 10^{-5}$~mbar) prior to annealing (labelled as Vacuum), and (c) after one hour of annealing at 400~K (in vacuum) (labelled as Vac. Anneal). The annealing conditions were kept the same for all WSe$_2$ and MoS$_2$ devices reported here. Apart from the evident increase in $I_{sd}$ on annealing, typically to about $\sim 100$~nA at largest experimental $V_g$ (Fig.~1c) ($V_{sd}=100$~mV), we also note: (i) First, a shift of the turn-on voltage $V_{ON}$, indicated by the dashed vertical lines, towards negative values. This has been observed earlier in WS$_2$ [REF:~\cite{Andras_Kis_WS2_anneal}] and MoS$_2$ [REF:~\cite{Shi_MoS2_anneal}] channels, and implies desorption of gas and water dopants. (Here, $V_{ON}$ is calculated by extrapolating the $\sqrt{I_{sd}}$ vs $V_{g}$ plot in the ON state of the transistor and finding its intercept on the $V_{g}$ axis~\cite{Streetman_Banerjee}.) This makes $p$-type conduction in our WSe$_2$ channel inaccessible within the experimental range of $V_g$, although it allows a wide operating range in the $n$-doped regime. (ii) Second, nearly four decades of increase in the field-effect mobility ($\mu = (1/C_{ox})\partial\sigma/\partial V_g$, where $C_{ox}$ is the gate capacitance per unit area) (Fig.~1d) to $\mu \approx 0.5 - 5$~cm$^2$/V.s, in agreement with previous reports~\cite{Andras_Kis_WS2_anneal}. (iii) Third, linear and symmetric $I_{sd} - V_{sd}$ characteristics at
\begin{figure*}[t]
\includegraphics[width=1\linewidth]{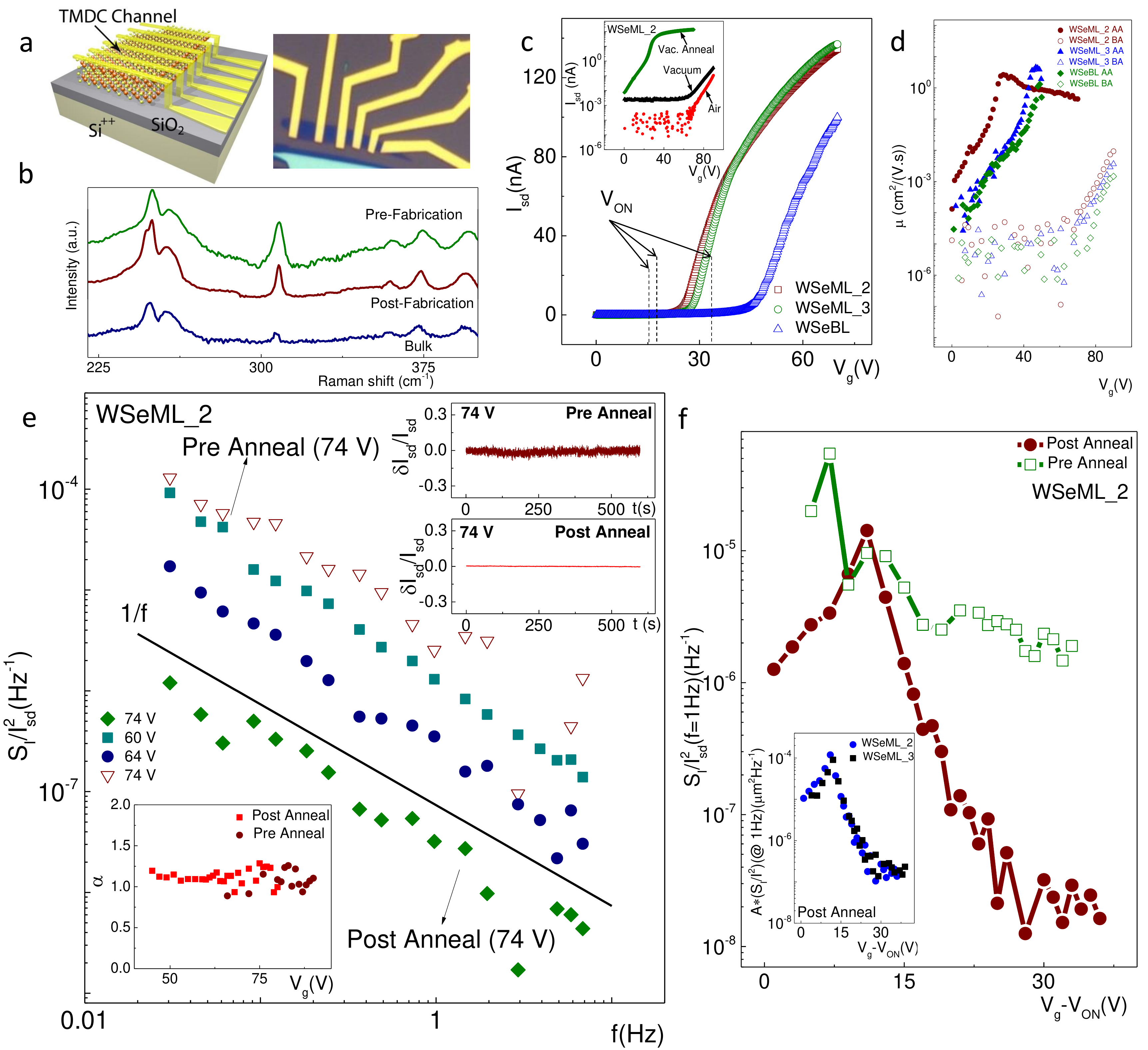}
\caption{Device structure and transport characterization. (a) Device Schematic and optical micrograph of a typical WSe$_2$ FET. (b) Raman Spectra for bulk WSe$_2$, WSe$_2$ flakes after exfoliation (Pre-fabrication) and following device processing (Post-fabrication). (c) Transfer charateristics of three WSe$_2$ FETs after annealing. The dashed lines indicate the turn-on voltage (see text). (Inset) A comparison of the transfer characterisitcs in air (labelled `Air'), in vacuum before annealing (labelled `Vacuum') and in vacuum after annealing (labelled `Vac. Anneal') for the device WSeML\_2. (d) Comparison of the field effect mobility before (BA) and after (AA) annealing curves for three devices WSeML\_2, WSeML\_3 and WSeBL. (e) Normalized power spectral density (PSD) of current fluctuation showing a $1/f^\alpha$-type frequency dependence. Lower inset: $V_g$-dependence of frequency exponent $\alpha$ both before and after annealing. A large reduction in the normalized time dependent current fluctuations is observed after annealing (upper inset). (f)  Comparison of normalized noise magnitude $S_I/I_{sd}^2$ for WSeML\_2 before and after annealing as a function of gate voltage $V_{g}$ measured from the turn-on voltage $V_{ON}$. Inset: Area-scaled normalized noise magnitude after annealing for two devices WSeML\_2 and WSeML\_3.}
\end{figure*}
low source-drain voltages (up to $|V_{sd}| \lesssim 100$~mV) (supplementary material), confirming the ohmic nature of the contacts. All electrical measurements were limited to the linear transport regime, where the fluctuations in $I_{sd}$ reflects that of $\sigma$. Thermal annealing in vacuum had a dramatic effect on the intrinsic low-frequency $1/f$-noise magnitude in WSe$_2$ FETs, particularly at large values of $V_g$ (high doping).
\begin{figure*}[t]
\includegraphics[width=1\linewidth]{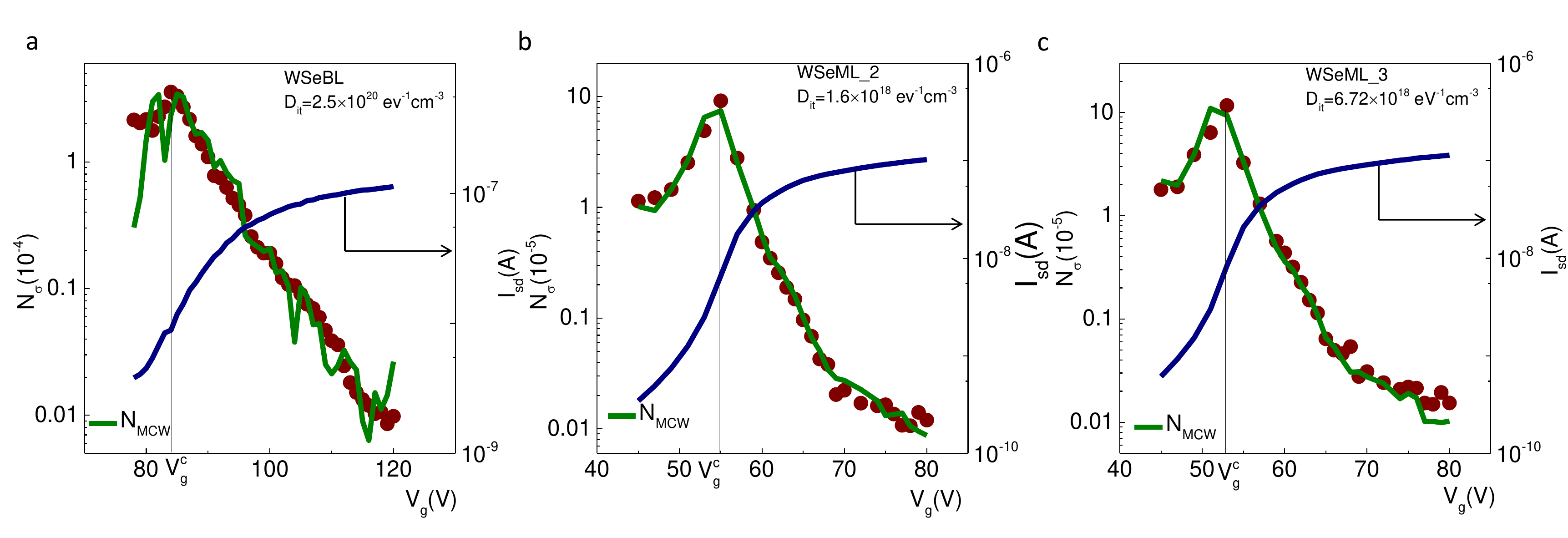}
\caption{Integrated noise and number fluctuation. Plots of integrated power spectral density (markers), calculated McWhorter noise $N_{MCW}$ (olive green colored line) using substrate trap density $D_{it}$ as fitting parameter (Equation~\ref{McWhorter}) and source drain current (navy blue line) for (a) WSeBL, (b) WSeML\_2, (c) WSeML\_3. The remarkable agreement between the calculated McWhorter noise with the experimental noise magnitude indicates channel carrier number fluctuations as the source of flicker noise in our WSe$_2$ channels.}
\end{figure*}
This is illustrated in the upper inset of Fig.~1e, which shows the normalized time-dependent current fluctuations $\delta I_{sd}/I_{sd}$ at the same gate bias before and after the annealing process. The overall shape of the power spectral density $S_I(f)$ of the fluctuations, which depends on frequency $f$ as $S_I(f) \propto 1/f^\alpha$ with $\alpha \approx 1$ (Fig.~1e), is insensitive to the annealing state or $V_g$ (lower inset of Fig.~1e). The main effect of annealing, however, lies in the $V_g$-dependence of the normalized noise magnitude $S_I/I_{sd}^2$, and shown in Fig.~1f with the data from device WSeML\_2. The pre-anneal $V_g$-dependence of $S_I/I_{sd}^2$ is weak with a monotonic decrease as $V_g$ is increased, whereas the variation in $S_I/I_{sd}^2$ becomes nonmonotonic after annealing, and can be smaller than the pre-anneal noise by more than two orders of magnitude at large $V_g$~\cite{Subhamoy_APL_Mat,Hersam_MoS2_Noise}. Quantitatively, the total measured noise can be decomposed into noise from the contact (normalized noise $s_{con}$) and the channel (normalized noise $s_{ch}$) regions, so that $S_I/I_{sd}^2 = \eta^2 s_{con} + (1-\eta)^2 s_{ch}$, with $\eta = G_{ch}/(G_{ch}+G_{con})$, where $G_{ch}$ and $G_{con}$ are the channel and contact conductances, respectively~\cite{Kaustuv_Banerjee_MoS2_noise}. Prior to annealing, $\eta \rightarrow 1$, and the contact noise is expected to dominate, while the post-annealing ($\eta < 10^{-4}$) noise is determined by the channel region. To confirm this we examined the dependence of the post-anneal normalized noise magnitude on the channel area ($A$) of two devices of different sizes from the same WSe$_2$ flake. As shown in the inset of Fig.~1f, $A\times S_I/I_{sd}^2$ from the devices collapse at all $V_g$, implying $S_I/I_{sd}^2 \propto 1/A$, which is expected when noise originates from the channel region~\cite{Saquib_PRB_2011}. (The $1/A$ dependence was not observed in the pre-annealed noise.) In the remainder of this paper we have confined our discussion only to the post-anneal noise behavior in the TMDC channels. Moreover, in order to analyze the noise data without referring to any specific frequency, we have computed the ``variance'' or the integrated power spectral density as $N_\sigma = \langle\delta \sigma^2\rangle/\langle \sigma\rangle^2 = (1/\langle I_{sd}\rangle^2)\int{S_I(f)df}$, where the integration was carried out within the experimental bandwidth.

In order to explore the microscopic mechanism of noise in WSe$_2$ FETs, we note the strong peak close to the steepest rise in the $I_{sd}-V_g$ transfer characteristics at a device-dependent gate voltage $V_g^c$ in all WSe$_2$ FETs (Fig.~2). This naturally suggests the noise from McWhorter carrier number fluctuations~\cite{Brini_MOSFET_noise}, where the fluctuation in $I_{sd}$ is due to trapping and detrapping of free carriers at the channel-oxide interface. This is equivalently represented by the flat-band gate voltage fluctuations, and thus resulting in current noise being proportional to the transconductance $g_m = \partial I_{sd}/\partial V_g$ of the channel. To establish this quantitatively, the noise magnitude can be written as~\cite{Kaustuv_Banerjee_MoS2_noise,Subhamoy_APL_Mat,Balandin_MoS2_noise,Junhong_Na_MoS2_noise,MoS2_noise_D_Sharma,Brini_MOSFET_noise},
\begin{equation}
\label{McWhorter}
N_{MCW} = \frac{\langle\delta I_{sd}^2\rangle}{\langle I_{sd}\rangle^2} = \frac{6.2e^2k_BT}{AC_{ox}^2\kappa}.D_{it}.\left(\frac{g_m}{I_{sd}}\right)^2
\end{equation}

\noindent where $D_{it} \approx 10^{18} - 10^{20}$~eV$^{-1}$cm$^{-3}$ is the density of trap states in SiO$_2$, and $\kappa \approx 10^{9}$~m$^{-1}$ is the inverse of electronic wave function decay scale inside the oxide. Using $D_{it}$ as the fitting parameter, and experimentally determined $g_m$, we have compared the $V_g$-dependence of the observed noise magnitude $N_\sigma$ and that expected from the McWhorter model (Equation~\ref{McWhorter}) for all WSe$_2$ FETs (Fig.~2). The excellent agreement with realistic range of values of $D_{it}$ for SiO$_2$ establishes the carrier number fluctuations at the channel-SiO$_2$ interface as the predominant source of noise in our WSe$_2$ FETs.

\begin{figure*}[t]
\includegraphics[width=1\linewidth]{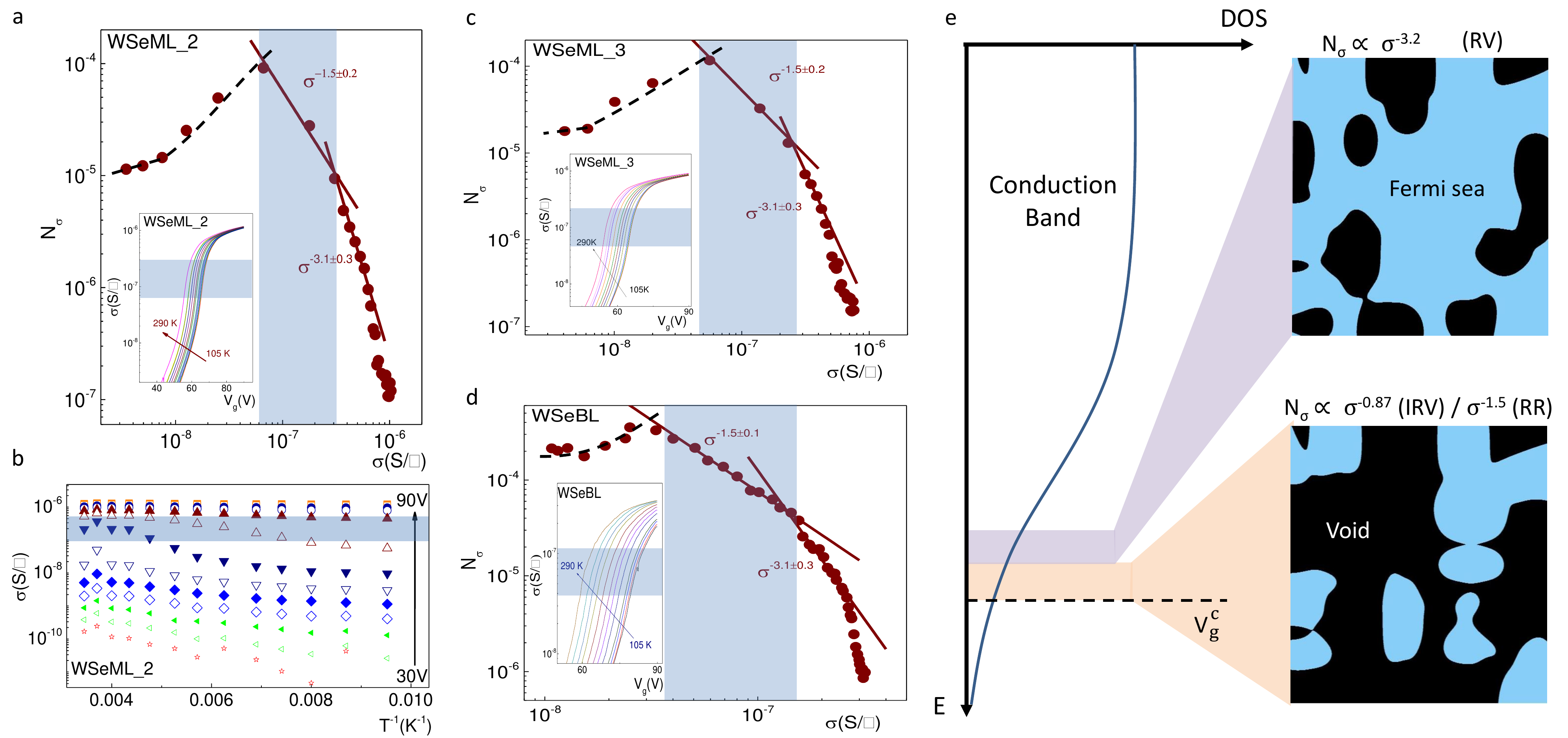}
\caption{Noise - conductivity scaling and percolation. Dependence of noise magnitude $N_\sigma$ on conductivity $\sigma$ and the gate voltage dependence of $\sigma$ (insets) for (a)WSeML\_2, (c)WSeML\_3, (d)WSeBL. Different noise scaling regimes are indicated by solid lines with respective exponents ($\nu$). The shaded region outlines the transition region with $\nu$ = 1.5. (b) Temperature dependence of $\sigma$ of WSeML\_2 indicating transition from localized to almost temperature independent transport. (e) Schematic of the ``island-and-sea" representation of charge distribution for different fermi level positions as a function of gate voltage. }
\end{figure*}

Subsequently, we have examined the dependence of $N_\sigma$ on the channel conductivity $\sigma$, which shows very similar trends for all three WSe$_2$ devices. In Fig.~3a, c and d, three distinct regions can be clearly identified: (i) Region~I: The low-conductivity regime at $V_g < V_g^c$, where $N_\sigma$ {\it increases} from a finite value with increasing $\sigma$ (dashed lines) to the maximum at $V_g^c$ which corresponds to $\sigma \simeq 3 - 5\times 10^{-8}$~S/$\square$, (ii) Region~II: at $V_g \gtrsim V_g^c$ extending over $\sigma \sim (0.5 \rightarrow 3)\times 10^{-7}$~S/$\square$ (indicated as the shaded area in Fig.~3a-d) where $N_\sigma \propto \sigma^{-(1.5\pm0.2)}$, and (iii) Region~III at $V_g \gg V_g^c$ and $\sigma \gtrsim 3\times 10^{-7}$~S/$\square$,  where $N_\sigma$ {\it decreases} rapidly as $N_\sigma \propto \sigma^{-(3.1\pm0.3)}$. The three regimes are characterized by distinct $T$-dependences of $\sigma$, as illustrated for WSeML\_2 in Fig.~3b, although other devices show very similar behavior (supplementary material)~\cite{Andras_Kis_mobility,Andras_Kis_WS2_anneal,Pablo_Bilayer_MoS2}. The weak $T$-dependence of $\sigma$ in Region~III signifies quasi-metallic electron transport. In Region~I and Region~II, {\it i.e.} around the switching transition ($\sigma \lesssim 10^{-7} -  10^{-8}$~S/$\square$), the insulating behavior becomes progressively stronger with decreasing $V_g$, although the signature of variable range hopping develops only at much lower $V_g$ {\it i.e.} stronger localization (see Fig.~3b and supplementary material for more detail), which makes identifying the transport mechanism at switching from $T$-dependence of $\sigma$ alone a very difficult task.

The exponents $\nu$ in $N_\sigma - \sigma$ scaling, however, provide a direct evidence of a percolative transport, with the noise peak signifying the percolation threshold at $V_g \approx V_g^c$~\cite{Tremblay_Percolation_Exponents,Kiss_Percolation_Noise,Percolation_Balberg}. Below the percolation threshold ($V_g < V_g^c$: Region~I), called the ``dielectric regime'', transport occurs by hopping or tunneling through disconnected metallic puddles~\cite{Percolation_Grimaldi}, and approaches a finite device-dependent magnitude at low $V_g$ away from the threshold.
In the critical (metallic) regime immediate to the percolation threshold ($V_g \gtrsim V_g^c$: Region~II), $\nu \approx 1.5$ corresponds to percolation in random resistor network (RR in Fig.~3(e)) with a broad distribution of intersite resistance~\cite{Kiss_Percolation_Noise}. On further increase in $V_g$, the crossover to $\nu \approx 3$ in Region~III marks the onset of continuum percolation with random voids, signifying transport in delocalized Fermi sea in the presence of insulating islands~\cite{Tremblay_Percolation_Exponents}.

Similar variation of $\nu$, and thereby the nature of percolative transport, has been observed earlier in a number of metal-insulator composites with varying fraction of the metallic component (see, for example, REF~\cite{Percolation_Balberg}). In our case, it allows a unique route to monitor the evolution of carrier distribution with gate voltage, or Fermi energy, especially at switching. Switching by percolation transition within the ``island-and-sea'' description in Fig.~3e, has been reported in several different classes of low-dimensional systems, including dilute 2D electron or hole gases in high mobility semiconductor heterostructures~\cite{S_Das_Sarma_Metal_Insulator_Transition,Yigal_Meir_Percolation,Aamir_PRB,Arindam_Spin_lattice_semiconductor_heterostructure}, silicon inversion layers~\cite{Si_inversion_layer_Shankar_Das_Sarma}, perovskite oxides~\cite{Anindita_PRB_2014}, graphene nanoribbons~\cite{Goldhaber_Disorder_graphene_nanoribbon,Graphene_Nanoribbon_Shankar}, or even MoS$_2$~\cite{C_T_Liang_disordered_MoS2,Ning_Wang_MIT_MoS2} and WS$_2$~\cite{Andras_Kis_WS2_anneal} FET devices. However the signatures of percolative transport in these systems are usually indirect, and depend on the observation of metal-insulator transition~\cite{S_Das_Sarma_Metal_Insulator_Transition,Yigal_Meir_Percolation}, or characteristic $I_{sd}-V_{sd}$ scaling~\cite{Middleton_Wingreeen_Percolation,Quantum_dots_Aamir_Srijit},
\begin{figure*}[t]
\includegraphics[width=1\linewidth]{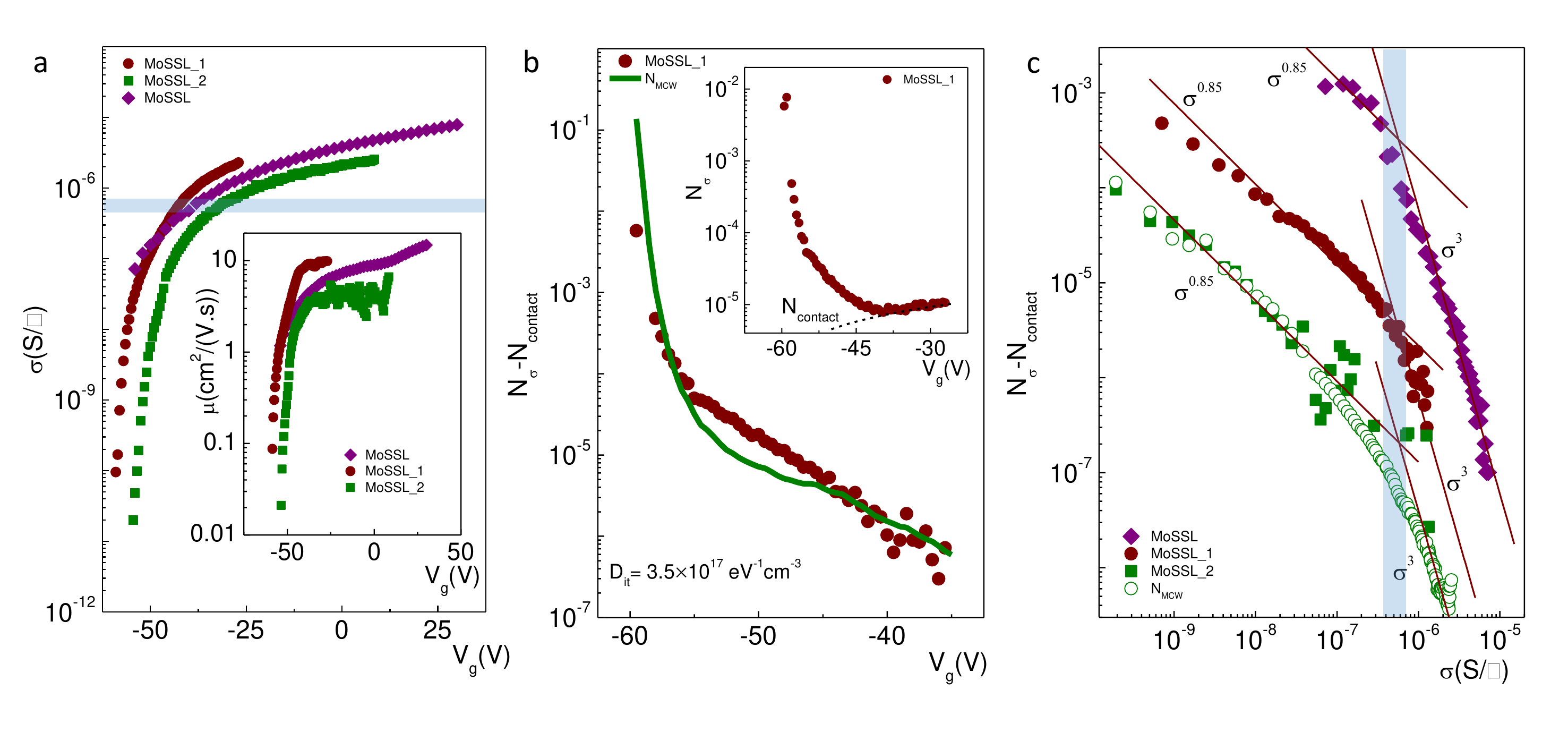}
\caption{Transport and Percolation Noise in MoS$_2$ FET. (a) Gate voltage dependence of conductivity ($\sigma$) and field effect mobility ($\mu$) (inset) for three MoS$_2$ devices MoSSL, MoSSL\_1 and MoSSL\_2. (b) Plot of normalised noise magnitude ($N_\sigma$) vs gate bias  ($V_g$) for MoSSL\_1 (inset). The dashed line indicates the noise contribution from the contacts. The main panel shows the gate voltage dependence of $N_\sigma$ corrected for contact noise ($N_{contact}$) and the calculated McWhorter noise (solid line) using trap density $D_{it}$ as a fitting parameter for device MoSSL\_1. (c) Scaling of $N_\sigma-N_{contact}$ with $\sigma$ for three MoS$_2$ devices showing characteristic percolative scaling exponents. The shaded region (in (a) and (c)) indicates the transition from the Inverted Random Void ($\nu \approx 0.85$) to Random Void ($\nu \approx 3$) mode of transport.}
\end{figure*}
which can also be limited to low temperatures and/or suffer from nonequilibrium effects. The noise-conductivity scaling constitutes a new probe to percolative transport in 2D TMDC materials, where the divergence of noise associates the switching transition of the FET to the percolation threshold. In order to verify the generality of the observed $N_\sigma-\sigma$ scaling in other classes of TMDC FETs, we have carried out noise experiment in similarly prepared backgated single layer MoS$_2$ FETs on SiO$_2$ substrate (see Table~I for details). Three different devices with field-effect mobility upto $\mu \sim 20$~cm$^2$/V.s (inset of \textbf{Fig.~4}a) were measured across the switching transition covering nearly five decades of change in conductivity (Fig.~4a). Fig.~4b (inset) illustrates the typical gate voltage dependence of $N_\sigma$ in one of the devices (MoSSL\_1), where we observe the noise magnitude to saturate at large doping ($V_g \gtrsim -40$~V). We have shown previously~\cite{Subhamoy_APL_Mat} that such a saturation arises from remnant effects of the contacts, and contributes additively to the total measured noise. Subtracting the contact noise $N_{contact}$ (dashed line in Fig.~4b inset), obtained by fitting the $V_g$-dependence of $N_\sigma$ at $V_g \geq -35$~V, we recover the overall agreement between the channel noise ($N_\sigma - N_{contact}$) and the calculated McWhorter noise $N_{MCW}$ (Equation~\ref{McWhorter}) over nearly five decades suggesting that the noise in the annealed MoS$_2$ FETs has a very similar microscopic origin as the WSe$_2$ devices.

Fig.~4c shows the scaling of $N_\sigma - N_{contact}$ with the channel conductivity $\sigma$ for all three devices (shifted vertically with respect to each other for clarity). The scaling behavior is very similar in all cases where $\nu \approx 0.85\pm0.15$ at low $\sigma$ and crosses over to the continuum 2D percolation with random void ($\nu \approx 3$) above a characteristic conductivity $\sigma_c \sim 5\times10^{-7}$~S/$\square$. The McWhorter noise, estimated from experimentally measured $g_m$ (shown in Fig.~4c for MoSSL\_2 with open circles) behaves in identical fashion over the entire range of $\sigma$ ensuring the consistency of the analysis. Interestingly, the magnitude of $\sigma_c$ is similar for WSe$_2$ and MoS$_2$ FETs in spite of different material and channel mobility. Importantly, MoS$_2$ channels only show a divergence of noise with decreasing $V_g$, and no peak is observed even down to the smallest measured $\sigma$ ($\sim 10^{-11}$~S/$\square$). This suggests that unlike WSe$_2$ FETs, conductivity below percolation threshold in MoS$_2$ FETs is too small to be measurable, possibly due to a predominantly long range nature of disorder. Hence the observed $\nu$ ($\approx 0.85$) at $\sigma < \sigma_c$ in MoS$_2$ is more likely to be due to continuum percolation in inverted random void (IRV, Fig.~3e) configuration,  rather than random resistor network. Nevertheless, the $N_\sigma-\sigma$ scaling in MoS$_2$ FETs with universal exponents suggests that percolative transport at switching could be generic in TMDC FETs, although the nature of percolation itself (discrete random resistor or continuum), is likely to depend on material details, channel mobility, disorder landscape, and other factors.

In conclusion, we have employed low-frequency $1/f$-noise in electrical conductivity to explore the nature of charge transport in FET devices with ultra thin TMDC channels, close to the switching transition at room temperature. The noise in both WSe$_2$ and MoS$_2$-based FETs arise from carrier number fluctuations, in agreement with previous experimental reports. Importantly, a unique scaling of the noise magnitude with channel conductivity in WSe$_2$ FETs provides direct evidence of percolative electron flow at switching which, in WSe$_2$ FETs, crosses over from the random resistor network to continuum percolation with increasing carrier density. Similar behavior in MoS$_2$ FETs suggest that percolative transport in inhomogeneous charge distribution at switching could be generic to other TMDC channels as well.

We acknowledge the Department of Science and Technology (DST) for a funded project.


\end{document}


\beginsupplement
\title{Supplementary material: Percolative switching in transition metal dichalcogenide field-effect transistors at room temperature}
\author{Tathagata Paul,$^1$ Subhamoy Ghatak,$^1$ and Arindam Ghosh$^1$}

\address{$^1$Department of Physics, Indian Institute of Science, Bangalore 560012, India}
\maketitle

\begin{table*}[h]
\centering
\renewcommand{\arraystretch}{1.3}
 \captionsetup{justification=centering}
\caption{Channel Length ($L$), Channel Width ($W$), details of electrical contact and layer number for the unannealed WSe$_2$ FET devices}
\label{tab:example}
\centering
\begin{tabular}{c|c|c|c|c}
    \hline
    Device Name & $L$ ($\mu$m) & $W$ ($\mu$m) & Contact Details & No. of layers\\
    \hline
    \hline

   WSeML\_1 & 0.5 & 3.4 & Au 40 nm & 4\\
    \hline

    WSeTL & 0.7 & 0.5 & Ti/Au 10nm/50nm & 3\\
    \hline
\end{tabular}
\end{table*}

\section{Experimental Details}
All devices were measured in a two probe geometry at a pressure of 1E$-$6 mbar inside a cryostat. For transfer characteristics source drain bias  to the device was supplied from LIASR830. The output current at the frequency of excitation was measured using the lock-in technique. Gate voltage was supplied using a Keithley 2400 DC sourcemeter. During I-V characteristics both gate voltage and source drain bias was applied using Keithley 2400 DC sourcemeters. For noise measurements, a bias voltage of 100 mV at a frequency of 220 Hz was applied to the sample. Time series data was obtained using NI DAC (NI USB-6210) at a rate of 1000 samples per second. The power spectrum was calculated by performing a fast fourier transform on this data. The Power spectral density follows $1/f$ law of flicker noise Fig: $2(b)$.
\newpage
\section{Bipolar Transport in WS\MakeLowercase{e}$_2$ FET\MakeLowercase{s}}
One of the major advantages of tungsten based TMDCs WS$_2$/WSe$_2$ over MoS$_2$ is the possibility of bipolar conduction owing to their smaller bandgap and reduced fermi level pinning. Indeed, bipolar transport in WSe$_2$ FETs has led to the recent demonstration of electrostatically tunable integrated photovoltaic and photodetection devices~\cite{Pablo_WSe2_pn_junc_Nat_Nano,Mueller_WSe2_solar_energy_conversion_Nat_Nano,Xiaodong_Xu_WSe2_Led,Andras_Kis_MoS2_photodetectors_Nat_Nano,Gomez_Photovoltaic,Mueller_Photovoltaic}, and also promises new designs of complementary digital logic by appropriate electrical contacting schemes~\cite{Kaustav_Banerjee_contacts_WSe2}. In our experiments, some unannealed WSe$_2$ FETs (details in Table S1) did show bipolar conduction (Fig. S1) although the ON-state current in these devices were too small for any meaningful noise experiments.

\begin{figure}[H]
\includegraphics[width=1\linewidth]{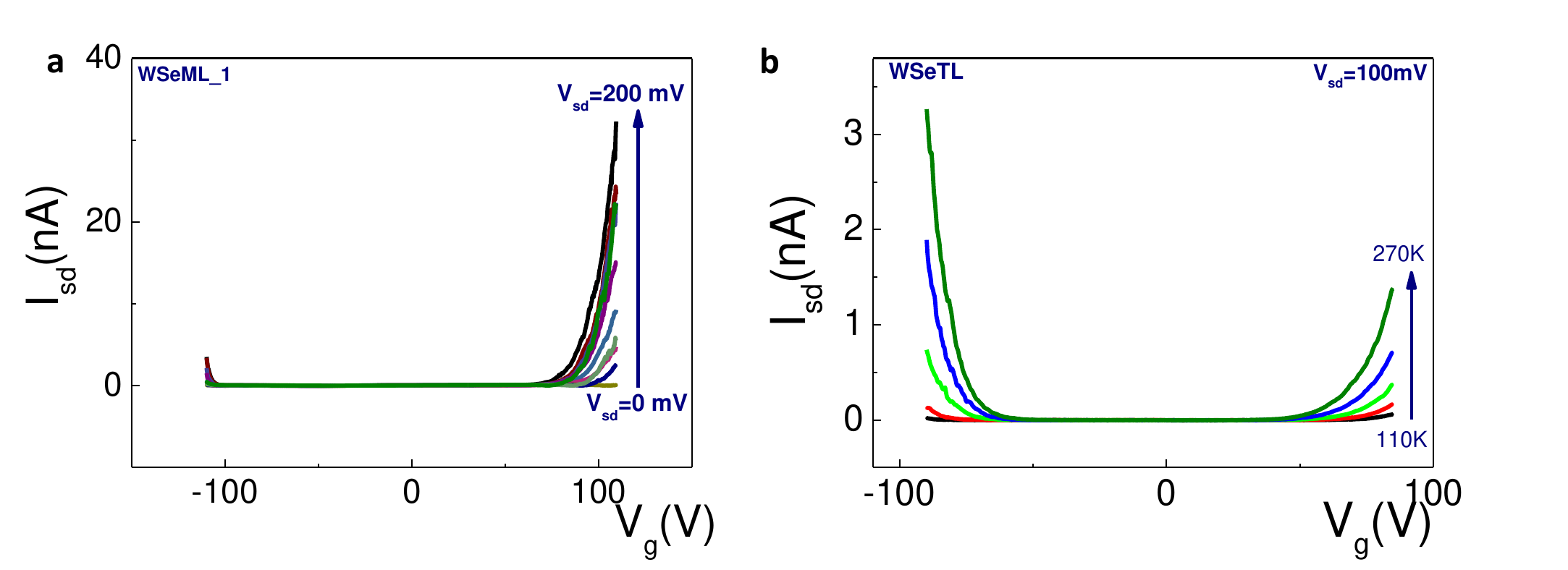}
\caption{(a) Transfer characteristics performed at 94.7~K for a range of source drain bias all showing bipolar channel conduction. The pinch off voltage for hole conduction was fairly high and gate oxide leakage prevented us from applying higher gate bias. Thus we see a small but prominent hole conduction. The mobility for electron and hole conduction was 0.6 and 0.1 cm$^2$/(V.s) respectively. (b) Temperature dependent transfer characteristics showing bipolar transport at all temperatures from 110~K to 270~K. Data was obtained at a fixed source drain bias of 100~mV. This strong bipolar nature of conduction is a proof of the comparatively low Fermi level pinnning in WSe$_2$ when compared to MoS$_2$. Maximum electron and hole mobilities for this device was 0.17 and 0.5 cm$^2$/(V.s) respectively. }
\end{figure}
\newpage
\section{Temperature dependence of conductivity ($\sigma$) and variable range hopping in WS\MakeLowercase{e}$_2$ FET\MakeLowercase{s}}
In Fig. S2, we have attempted to identify the microscopic transport mechanism, specifically the 2D variable range hopping or thermal activation. At low values of gate voltage ($V_g$) we observe an insulating behaviour which transforms to an almost temperature independent or slightly metallic behaviour at higher values of gate voltage. Neither two dimensional variable range hopping (Fig. S2) nor activated transport (shown for WSEML\_2 in Fig. 3b) can describe the $T$-dependence of conductivity ($\sigma$) adequately. However, at very low gate voltages ($V_g<30$~V),~$\ln\sigma$ tends to become linear with $T^{-0.33}$ (Fig. S2), probably indicating contributions from 2D variable range hopping.
\begin{figure}[H]
\includegraphics[width=1\linewidth]{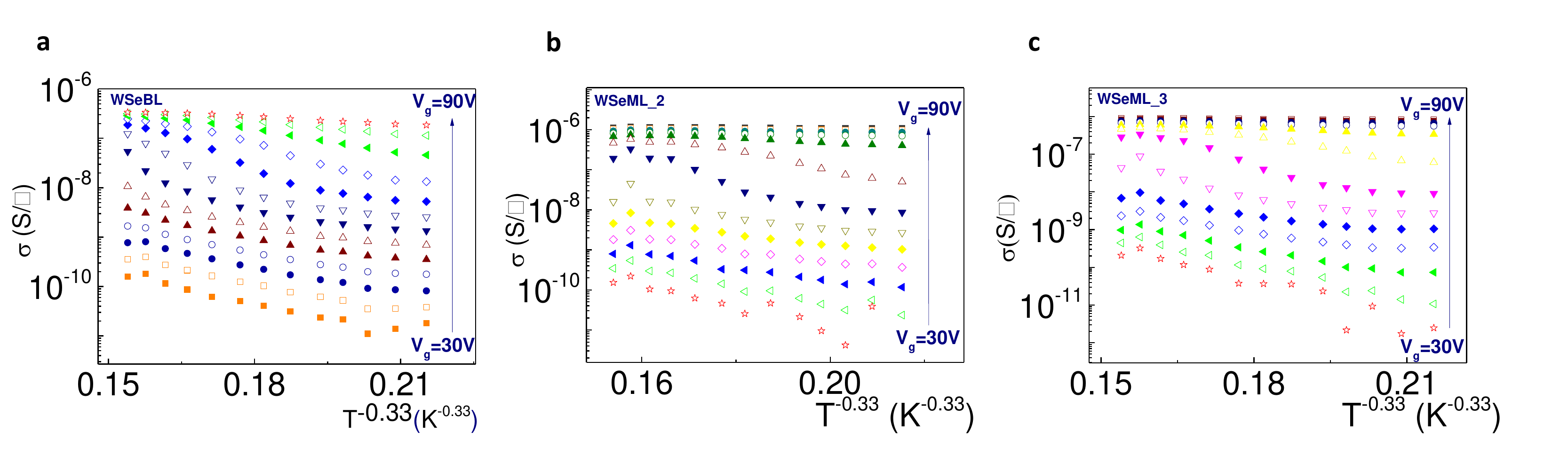}
\caption{(a),(b) and (c) Temperature dependence of conductivity for three separate WSe$_2$ FETs showing insulating behaviour  for lower gate voltages which evolves into an almost temperature independent transport for higher gate voltages. The transition occurs through a percolative mechanism of transport as  confirmed by the scaling of noise exponents as shown in the main body of the paper.}
\end{figure}
\newpage
\section{Effect of annealing on transfer characteristics of WS\MakeLowercase{e}$_2$ FET\MakeLowercase{s}}
As stated in the main text, annealing leads to a significant increase in the drain current $I_{sd}$, shifting of the turn-on voltage $V_{ON}$ towards negative values and increase in mobility by almost four decades for WSe$_2$ FETs. These features were seen for all our annealed WSe$_2$ devices and is shown here for WSeBL (Fig S3(a)) and WSeML\_3 (Fig S3(b)).

\begin{figure}[H]
\includegraphics[width=1\linewidth]{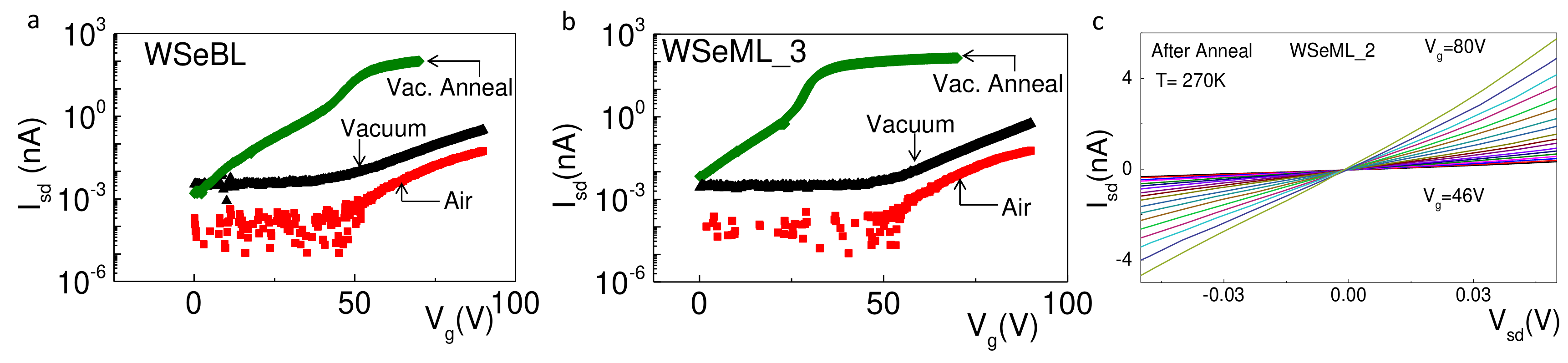}
\caption{A comparison of the transfer characteristics in air (labelled Air), in vacuum before annealing (labelled Vacuum) and in vacuum after annealing (labelled Vac. Anneal) for the device WSeBL (a) and WSeML\_3 (b). (c) $I_{sd}-V_{sd}$ characteristics of WSeML\_2 showing linear dependence at low bias.}
\end{figure}

\section{Raman Spectroscopy and AFM characterization of M\MakeLowercase{o}S$_2$}
\begin{figure}[H]
\includegraphics[width=1\linewidth]{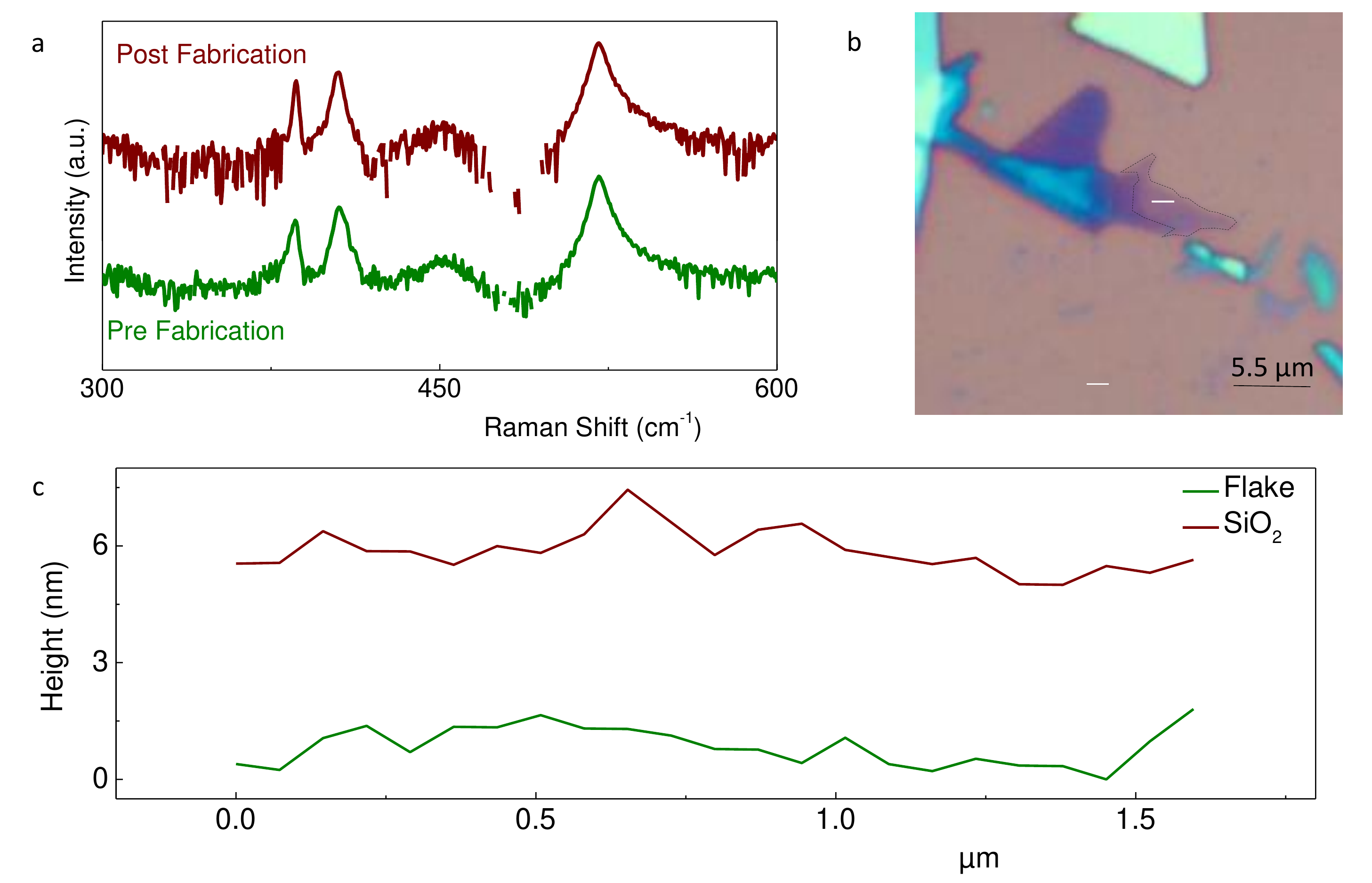}
\caption{(a) Raman characterization for MoS$_2$ flakes before and after device fabrication. Raman data for MoS$_2$ shows peaks for $E_1^{2g}$ mode at $\approx 385$~cm$^{-1}$ and $A_{1g}$ mode at $\approx 405$~cm$^{-1}$. The raman peak at $\approx 520$~cm$^{-1}$ is due to the silicon substrate. The position of the peaks match perfectly with previously reported data~\cite{Raman_Mos2}. This leads us to believe that the crystal structure is not adversely affected by the fabrication process. We used a laser wavelength of $532$~nm for our measurements. (b) Optical micrograph of an exfoliated MoS$_2$ flake. The white lines indicate the paths along which the height profile (panel (c)) for the MoS$_2$ and SiO$_2$ surfaces (shifted vertically for clarity) have been measured. The surface roughness in the channel region is $\approx 0.604$~nm whereas that on bare SiO$_2$ is $\approx 0.65$~nm. The similarity of these values indicate the conformity of the flakes to the SiO$_2$ surface. This percludes the presence of ruptures or other sharp structures in the channel region which might affect the transport properties adversely. Surface roughness was measured by computing the standard deviation of the height data in a given region of the channel or SiO$_2$ surface.}
\end{figure}

\section{AFM characterization of WS\MakeLowercase{e}$_2$}
\begin{figure}[H]
\includegraphics[width=1\linewidth]{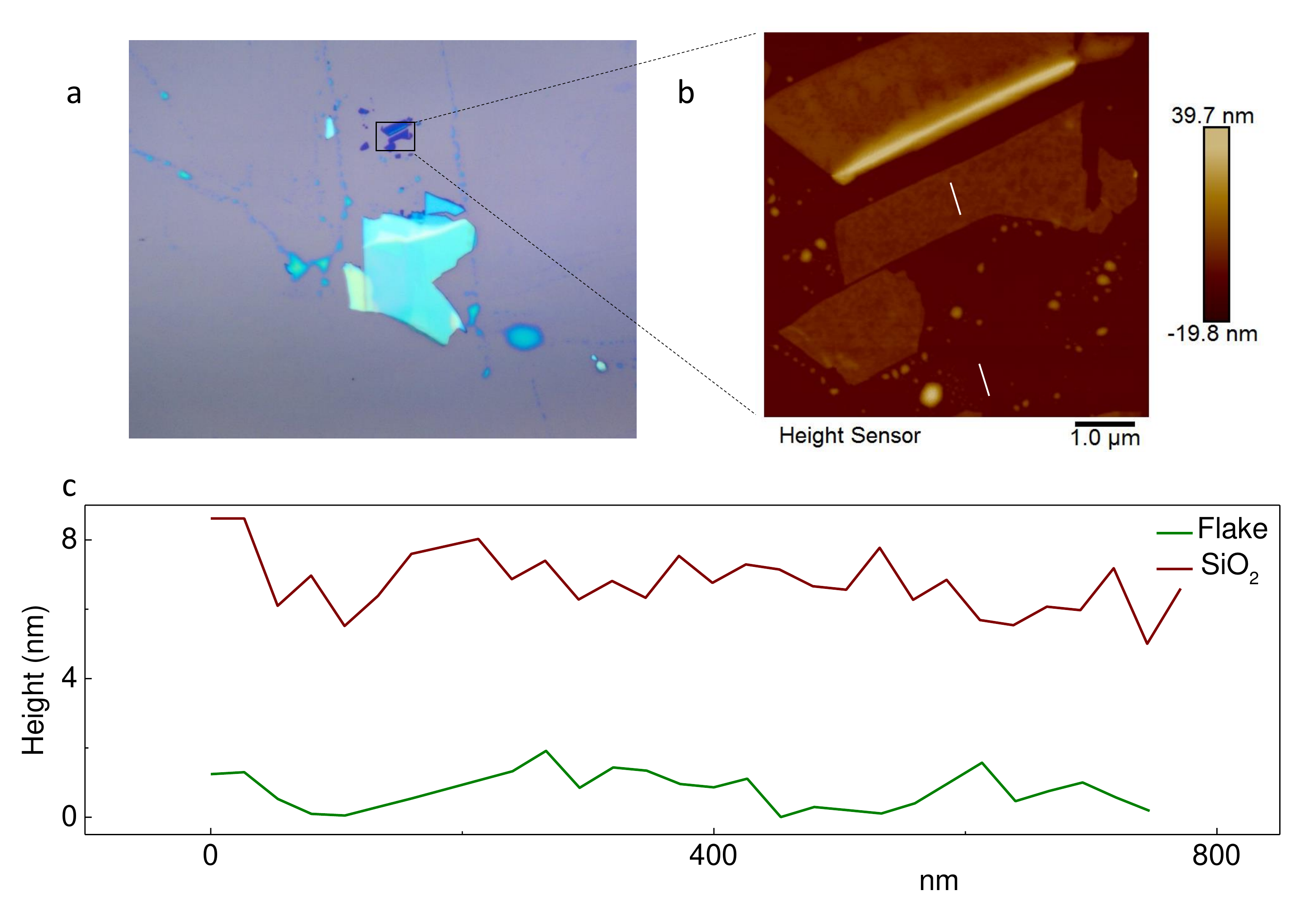}
\caption{Optical micrograph (a) and a zoomed in AFM image (b) of a typical exfoliated WSe$_2$ flake. The height profile (c) (shifted vertically for clarity) has been measured along the paths (white lines) shown in panel (b). Roughness for both the flake and the SiO$_2$ surface was calculated to be $\approx 1$~nm. Hence, the WSe$_2$ flakes as in the case of MoS$_2$ ones (Fig~S4) conform to the SiO$_2$ surface.}
\end{figure}
%
